# The background seismicity rate of the Greek territory, considered as a unified seismogenic area, for the period 1964 – 2010. Is Greece in course for a large seismic event?

Thanassoulas[1], C., Klentos[2], V., Verveniotis[3], G., Zymaris[4], N.

1. Retired from the Institute for Geology and Mineral Exploration (IGME), Geophysical Department, Athens, Greece.
   e-mail: thandin@otenet.gr - URL: www.earthquakeprediction.gr

2. Athens Water Supply & Sewerage Company (EYDAP),
   e-mail: klenvas@mycosmos.gr - URL: www.earthquakeprediction.gr

3. Ass. Director, Physics Teacher at 2nd Senior High School of Pyrgos, Greece.
   e-mail: gver36@otenet.gr - URL: www.earthquakeprediction.gr

4. Retired, Electronic Engineer.

**Abstract**

The analysis of the earthquakes catalog of Greece for the period 1964 to 2010 indicates a continuously increasing no. of seismic events. A detailed determination of the cumulative seismic energy release, for specific earthquake magnitudes, revealed that for magnitudes up to Ms = 5.0R the Greek territory, considered as a unified seismogenic area, is at a state of accelerating deformation for almost the last 10 years. The background seismicity rate, considering constant cumulative seismic energy release for the entire study period, corresponds to a magnitude of Ms = 5.5R, while for earthquakes of larger magnitudes the Greek seismogenic area behaves irregularly. The results are compared to the different phases (Mjachkin et al, 1975) a seismogenic area undergoes before a large seismic event. From this comparison it is concluded that the Greek territory is probably in the course for a large seismic event in the years to come.

**Key words:** background seismicity rate, accelerating deformation, Greece, cumulative seismic energy release, earthquake magnitude.

## 1. Introduction.

Seismicity, generally, consists of two parts: (1) the earthquakes that occur by themselves and are due to the strain accumulated by tectonic forces and unknown redistributions of stress and (2) the earthquakes that depend on others, such as aftershocks and foreshocks, swarms, and multiplets that are referred as clusters. In many studies, the assumption is made, often implicitly, that the rate of the non-clustered seismicity is stationary Poissonian with time (Wyss et al. 2000).

The rate of seismic events which occur by themselves in a region and is considered as normal is referred as background seismicity rate. The separation of the two parts (background and clustered) of seismicity ranges from easy to impossible, but in most cases a satisfactory, although not perfect, separation can be reached.

The major problems encountered in the determination of background seismicity rates are dependent events and detection or reporting changes. Dependent events, conceptually, are those that occur because of stress or strength changes caused by a previous earthquake(s). Many methodologies that resolve the problem of separation of the depended from the independent seismic events have been proposed i.e. Gardner and Knopoff, 1974; Hadley and Cavit, 1982; Savage (1972); McNally (1976).

Reporting and detection changes are unavoidable in the earthquake catalogs since these catalogs are the result of a large amount of work by a group of individuals and machines. Changes in the way this system detects and reports earthquakes are unavoidable simply because of the size of the system (Habermann and Wyss, 1984). Methods that deal with this problem have been presented by Habermann (1983).

In this work the background seismicity rate of the Greek territory will be studied in a non-stochastic methodology. Specifically, it will be analyzed in terms of seismic energy release (Thanassoulas, 2007). More over, the regional Greek territory will be considered as a unified seismogenic area. The study will mainly cover the time period from 1964 to 2010.

## 2. The Greek seismogenic territory.

Greece, being located at the place where the African, Anatolian and the Appoulian plates collide, is characterized by intense seismicity. Actually, it is the most seismically active country in Europe and among the most seismically active countries all over the world. Its high degree of seismicity attracted the research interest of seismologists who studied in detail its seismological and geotectonic regime. Some characteristic results are presented as follows. De Bremaecker et al. (1982) presented a finite element network with boundary conditions velocities, regarding the movement of the European plate for the Greek regional seismogenic area while the kinematics of the Greek territory tectonic plates were studied by Papazachos et al. (1996) too.

The kinematics, of the Greek territory plates, exhibits an impact on the observed seismicity of the same regional area. The latter is demonstrated in the study of the earthquakes which occurred from 600BC to 1986 presented by Papazachos (1988). Thus, the Greek territory has been divided into narrow seismically active zones while Hanus and Vanek (1993) mapped in a different approach the Greek seismically active zones too.

A quite different approach was followed by Thanassoulas (1998, 2007) in order to study the deeper regional tectonics of the Greek territory. The analysis of the Greek area gravity field and specifically the mapping of its horizontal gradient revealed the lithospheric fracture zones and faults where mostly the large (Ms>6.0R) earthquakes take place. The results of this analysis revealed the large degree of localized coincidence of the large earthquakes to the identified deep lithospheric fracture zones and faults. Adopting the notion of the lithosphere as being an open physical system that absorbs strain energy and releases it in the form of earthquakes, it was possible to calculate the stored seismic energy (seismic potential map) that could be released at any place in the Greek territory by manipulating the past seismic history of the Greek territory (Thanassoulas and Klentos, 2003, 2008, 2010).

Furthermore, as it is shown by the circular feature of the seismic potential map, the Aegean micro-plate that accounts for the central part of the Greek territory, rotates counterclockwise while Its rotation period has been calculated as T = 113.6my (Thanassoulas, 2007).

It is clear that the observed seismicity of the Greek territory closely depends on the kinematics of the regional area and therefore on the spatial distribution of the tectonic stress at any time. Some basic observations of the Greek seismicity based on simple statistical methods will be presented in the next section.

**2.1 Greece as a unit regional seismogenic area.**

The main kinematics mechanism of the Greek territory is its South-West drift combined to a CCW rotation. Therefore, the spatial distribution of the stress-field observed in the regional area is controlled by the present kinematics mechanism and consequently the generated seismicity is controlled too. Thus, the Greek territory is considered as a unit regional seismogenic area which is controlled by a single driving mechanism. Having adopted this working model, the earthquakes that occurred during the period 1901 – 2010 will be processed. The earthquake catalog of the National Observatory of Athens (NOA) of the Geodynamic Institute of Greece, which will be used for the study period, due to reporting and detection changes can be distinguished into two parts. The first one (1901 – 1964) covers the period when a regional seismological network was absent, and as a result earthquakes below a magnitude of 3.5 – 4.0R are missing. On the contrary, the second part of the earthquake catalog (1965 – 2010), when the National Seismological Observatory Network of Greece was put in operation on 1965, the detection threshold of the earthquakes magnitude was lowered to almost 1.2R. Since the Greek regional seismogenic area is considered as a unified one and taking into account that the energy release will be studied instead of discrete seismic events, de-clustering was not applied on the earthquake catalog. Therefore, only the 1964 – 2010 period part of the earthquake catalog was used for a brief presentation in figures (1, 2) while the minimum threshold earthquake magnitude level was kept at Ms = 3.0R.

The left graph of figure (1) presents the no. of seismic events per month, regardless their magnitude. The magnitude taken into account ranges from 3.0R to 7.5R. It is worth noting the increase of the seismic events that takes place during almost the last 15 years. The right graph of figure (1) presents the corresponding cumulative seismic energy release. In this graph too, increase of seismic energy release is observed during almost the last 10 years.

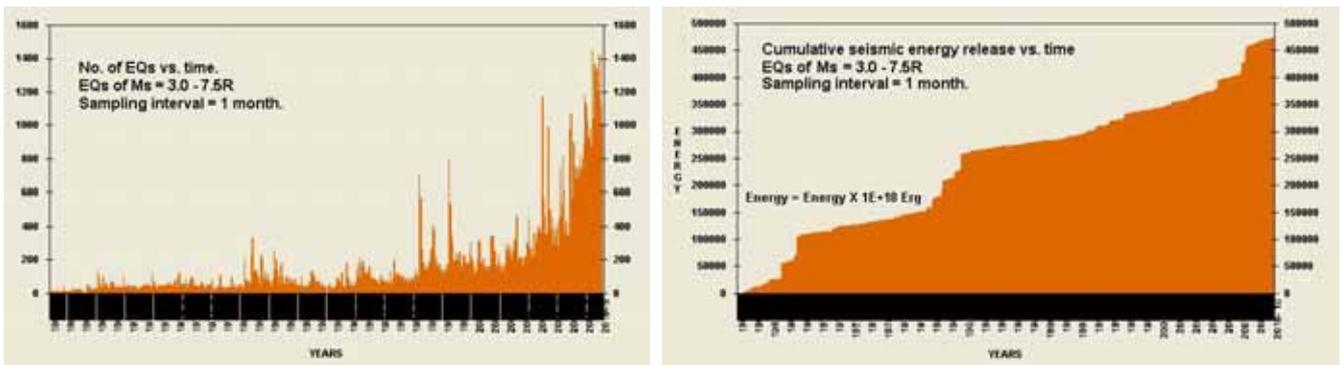

**Fig. 1. No. of seismic events per month (left) and the corresponding cumulative seismic energy release (right) for the period 1964 – 2010.**

The noisy character of both graphs of figure (1) can be eliminated by choosing a longer sampling interval. In the following figure (2) the same data were processed with a sampling interval of 6 months.

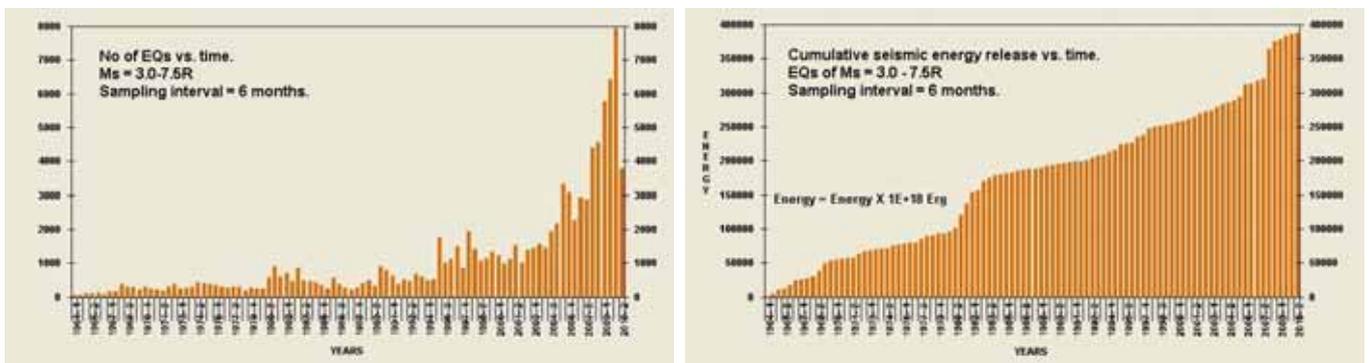

**Fig. 2. No. of seismic events per 6 months (left) and the corresponding cumulative seismic energy release (right) for the period 1964 – 2010.**

Time periods of intense seismic activity are clearly shown by characteristic increase steps of the cumulative seismic energy release graph (fig. 2, right).

Although the no. of seismic events as a function of time (in a seismogenic area) is important, the clustering of the earthquakes in the magnitude space is important for the seismotectonic characterization of a seismogenic area. The latter is expressed by the well-known parameter "b" of the Gutenberg – Richter relation:

$$\log(N) = a - bM$$

The occurrence frequency of the seismic events of the Greek territory for the period 1901 – 2010 vs. magnitude is presented in the following figure (3). The used magnitude sampling interval is 0.2R and the earthquake catalog spans from 1901 to 2010 so that all seismic events will be taken into account even if the absence of small earthquakes of the period 1901 – 1964 slightly modifies the corresponding graph.

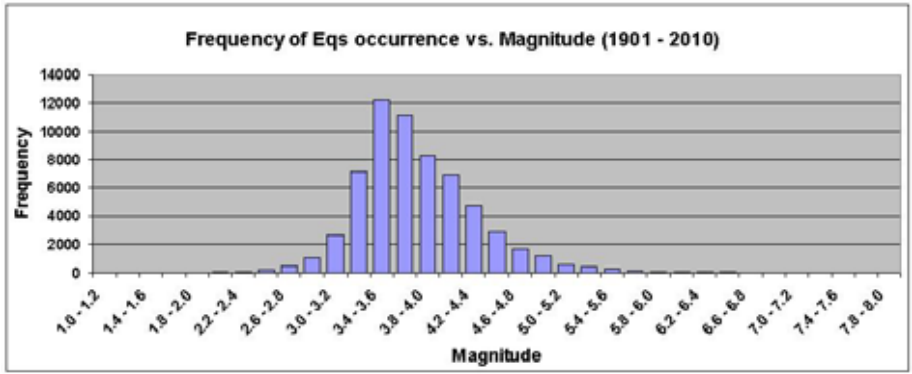

Fig. 3. Earthquake occurrence frequency vs. magnitude.

It is evident that figure (3) represents a typical stationary in time Poissonian distribution which complies with what is expected for a large seismogenic volume where so many separate tectonic regions contribute to its overall seismicity.

From figure (3) it is shown that the highest occurrence of seismic events is observed for the magnitudes 3.4R<Ms<3.6R. If a fixed average magnitude value of Ms = 3.5R is chosen then it is possible to compile the spatial distribution of the most frequent seismic events. That operation resembles the one of a very sharp band-pass filter. The compiled map is presented in figure (4).

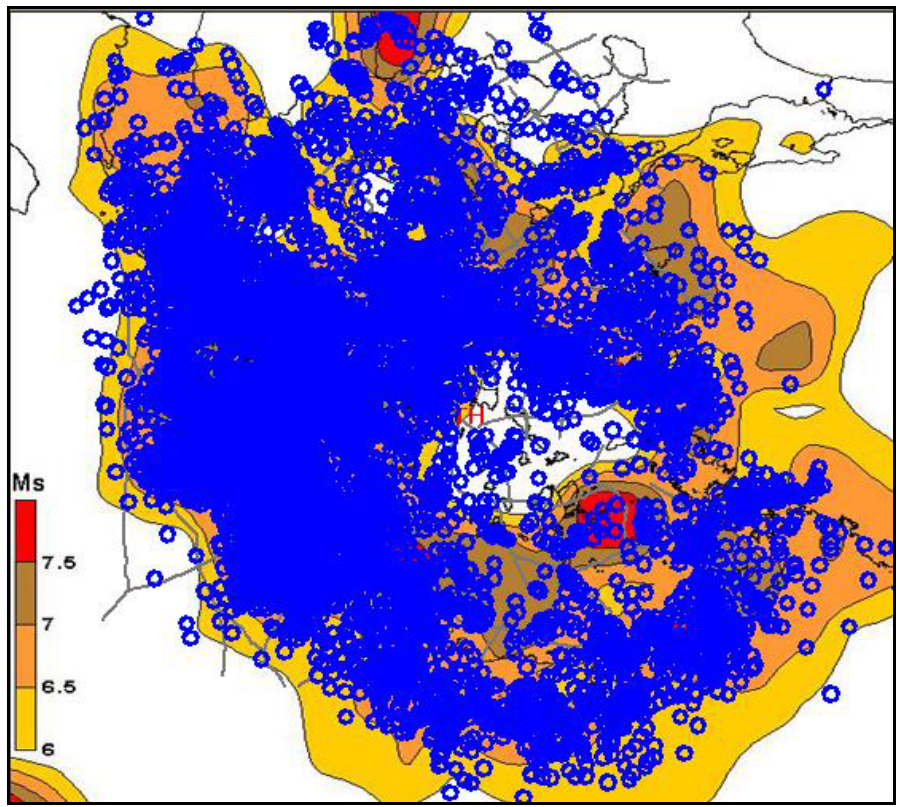

Fig. 4. Spatial distribution of seismic events with magnitude Ms = 3.5R

What is interesting in figure (4) is the fact that the earthquake locations comply to the seismic potential of the Greek territory (year 2010) where earthquakes larger than Ms = 6.0R are expected and further more their circular spatial distribution character which complies with the CCW rotation of the Aegean plate.

Furthermore, the data of figure (3) can be used to evaluate the corresponding "b" parameter value for the regional Greek seismogenic area. Taking the logarithmic value of the occurrence frequency of the seismic events, figure (3) is transformed into figure (5).

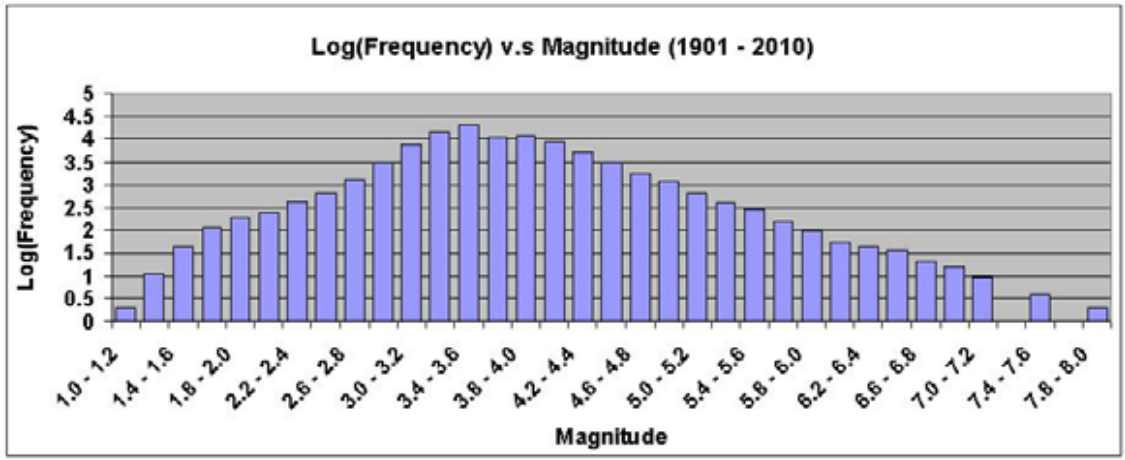

**Fig. 5. Log(frequency) of earthquake occurrence vs. magnitude.**

And from the right part of figure (5) the "b" parameter value is determined as follows and presented in figure (6):

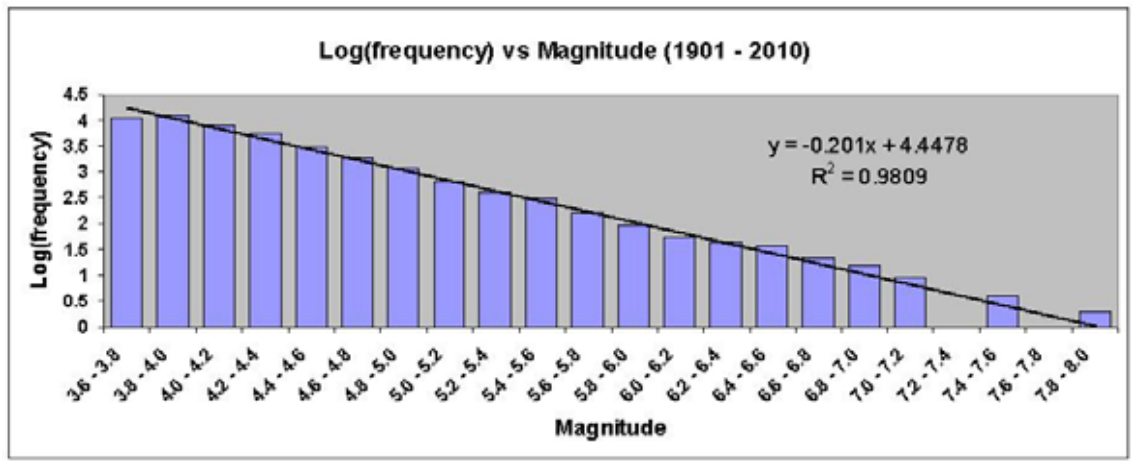

**Fig. 6. Parameter "b" value determination.**

In detail:  "b" = scale length x magnitude interval x 0.201 = 0.88

$$b = 0.88$$

Earlier determinations of the "b" parameter value for Greece presented by Papazachos (1988) have resulted as follows:

for South – Western Greece:   "b" = 1.0

for Central Greece:            "b" = 0.8

for North – Eastern Greece:    "b" = 0.6

By combining all these three values by averaging into one then the parameter "b" average value can be calculated ("b" average = 0.8) which value is very close to the one ("b" = 0.88) already calculated considering the entire Greek territory as a unified seismogenic area.

A more representative picture of the seismicity spatial distribution is obtained by using "sharp band-pass" filters for specific earthquake magnitudes and by compiling the corresponding spatial distribution maps. Following (fig. 7, 8, 9) are shown the compiled maps at an interval of one (1) unit of the Richter scale for the range of Ms = 3.0 – 8.0 R.

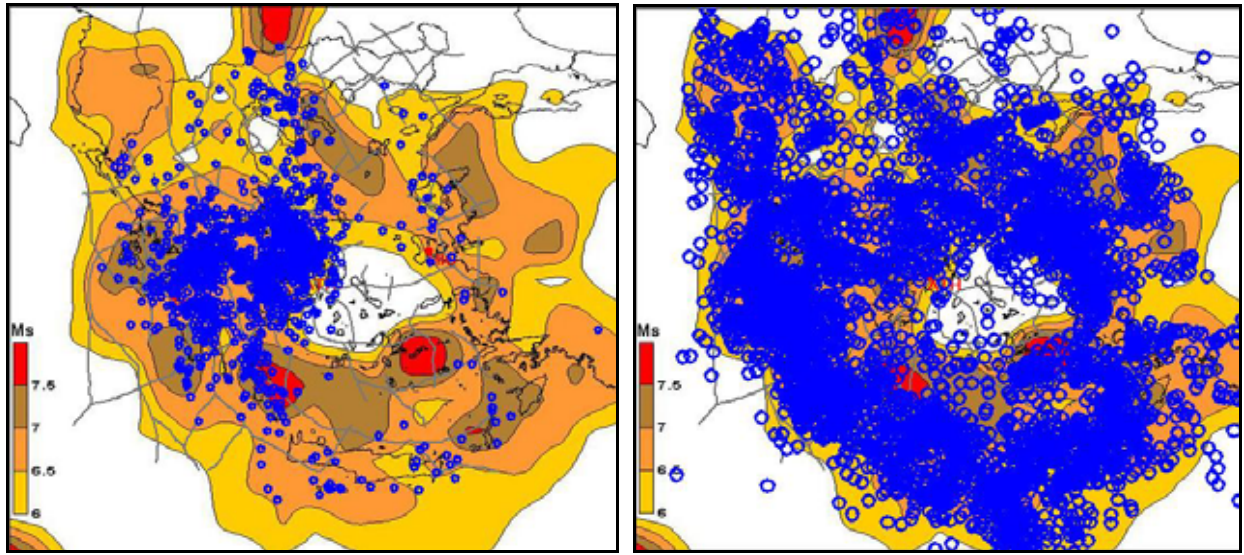

**Fig. 7.** Spatial distribution of earthquakes of Ms = 3.0R (left) and Ms = 4.0R (right) for the period 1901 – 2010.

Since magnitudes less than Ms = 3.0R were not registered before 1964 the left map of figure (7) represents the seismicity rate for the period 1964 – 2010. Two characteristic features are very interesting in this map. The first is that the seismicity of this level (Ms = 3.0R) is clustered in two rather narrow regions. The latter could be explained by: a. these two areas are the most seismically active of the Greek territory or b. if this type of seismicity increases with time then a rather large seismic event slowly in time evolves in this area. The second feature is that even small earthquakes are circularly clustered at the Aegean plate so they suggest indirectly its rotation. In the right map of figure (7) the circular pattern that characterizes the spatial distribution of the earthquakes of Ms = 4.0R is more obvious while it complies quite well with the corresponding background seismic potential map of Ms > 6.0R.

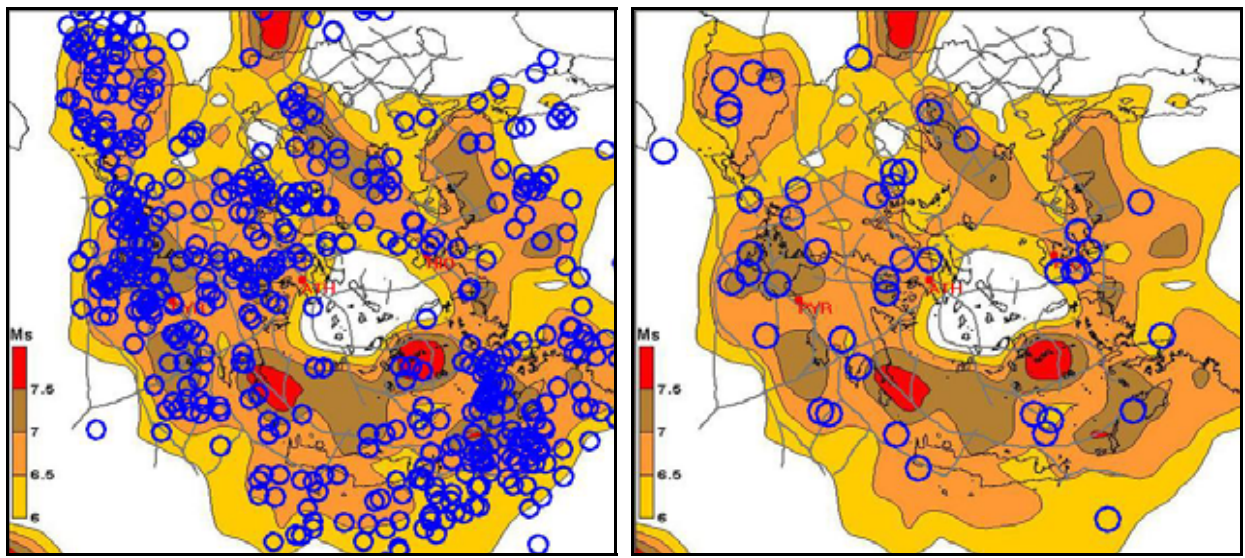

**Fig. 8.** Spatial distribution of earthquakes of Ms = 5.0R (left) and Ms = 6.0R (right) for the period 1901 – 2010.

The Ms = 5.0R magnitude earthquakes (fig. 8 left) although are fewer compared to the previous magnitude level (Ms = 4.0R) they are still circularly spatial distributed following the rotational character of the Aegean plate and comply with the seismic potential map of Greece. At the next magnitude level of Ms = 6.0R (fig. 8 right) the circular character of the spatial distribution almost vanishes and a rather random one prevails. Even in this case the vast majority of these earthquakes are located in the area of the seismic potential map where earthquakes with magnitude Ms > 6.5R are expected in the future.

Finally, the earthquakes of magnitude Ms = 7.0R and 8.0R are considered. Their spatial distribution is presented in the following figure (9).

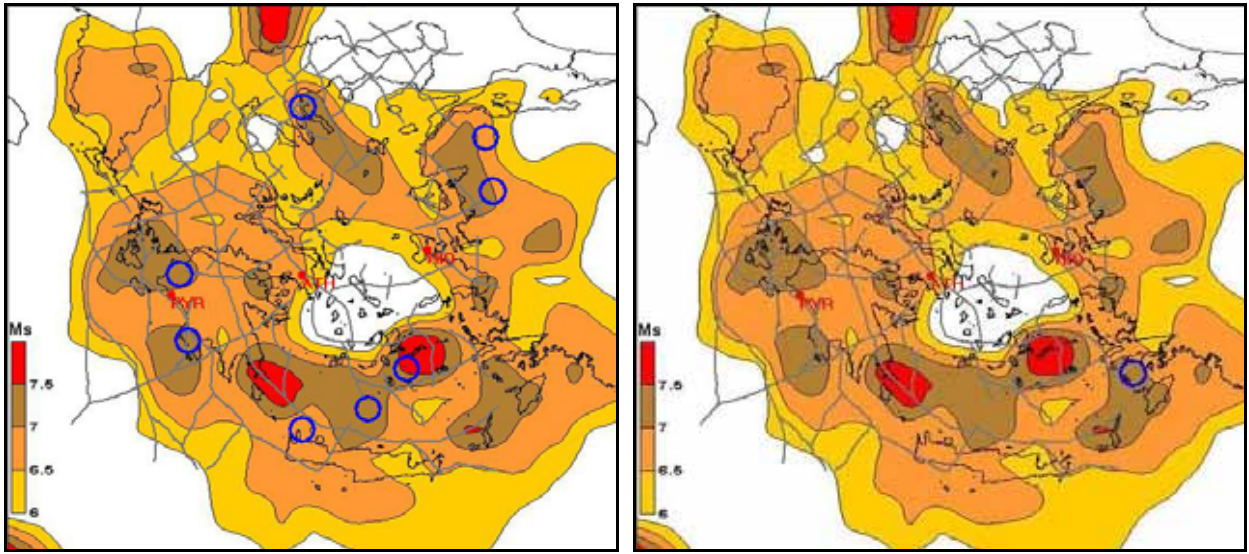

**Fig. 9. Spatial distribution of earthquakes of Ms = 7.0R (left) and Ms = 8.0R (right) for the period 1901 – 2010.**

In figure (9) left, although the earthquakes of magnitude Ms = 7.0R are very few (only 8), six (6) out of (8) took place in the areas where it is suggested that earthquakes of Ms > 7.0R will take place in the future, one (1) "failed" being located at an area of expected magnitude of 6.5R < Ms < 7.0R and one (1) is located marginally at the area of expected earthquakes of Ms > 7.0R. What is more interesting in figure (9) left is the fact that large and obviously catastrophic earthquakes took place in narrow defined seismic areas of large stored seismic energy thus validating the used methodology for the seismic potential map compilation and more importantly indicate in detail the narrow zones which are prone to large seismic events in the future. In figure (9) right, only one (1) earthquake of Ms = 8.0R was registered in the study period and is shown. This single seismic event fosters the results presented in figure (9) left.

**2.2. Time analysis in year's sample at 10 years intervals.**

So far the stationary seismicity as well as the tectonic character of the Greek seismogenic territory has been presented. But the earthquakes occurrence is a physical event that depends on time varying parameters of the lithosphere. Stress-load and strain due to compressional or extensional forces present in the lithosphere are the most important. Therefore, it is worth to study the evolution of earthquakes in time. To this end the magnitude of earthquakes which occurred within a sample year will be plot as a function of occurrence time for sample years at 10 years interval. Since the change of the stress-load conditions in the lithosphere is very slow it requires a rather long sampling interval so that an observable difference on the seismicity can be detected. Therefore, a ten-year interval was chosen and the referred analysis was applied for the years: 1965, 1975, 1985, 1995, 2005 and 2010 so that the last five-year's seismicity data could be included. The obtained results are presented in detail in the following figures (10) to (15). Each graph is accompanied by the magnitude frequency distribution at 0.5R intervals.

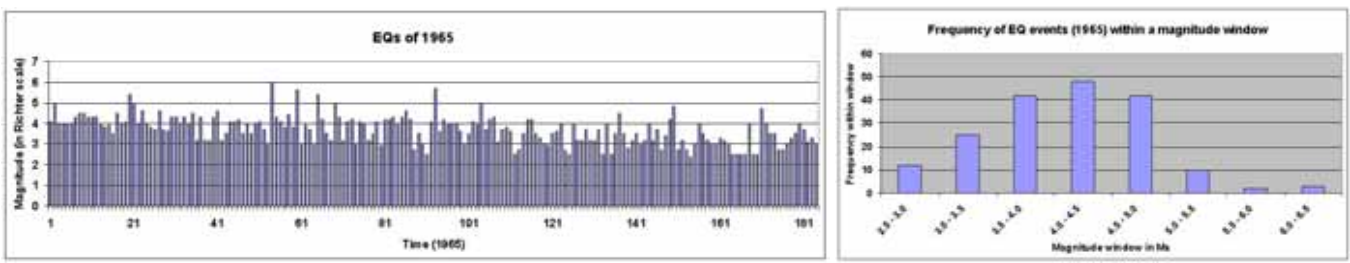

**Fig. 10. Magnitude as a function of occurrence time (left) and frequency distribution of magnitudes at 0.5R intervals (right) for the year 1965.**

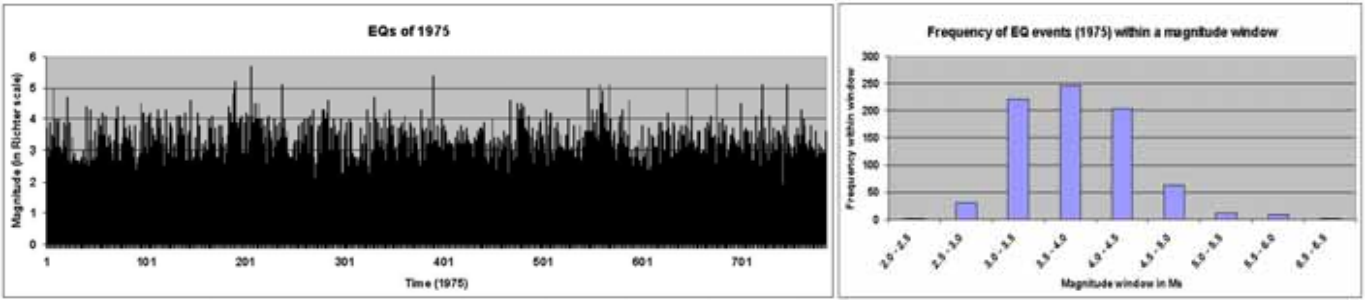

**Fig. 11. Magnitude as a function of occurrence time (left) and frequency distribution of magnitudes at 0.5R intervals (right) for the year 1975.**

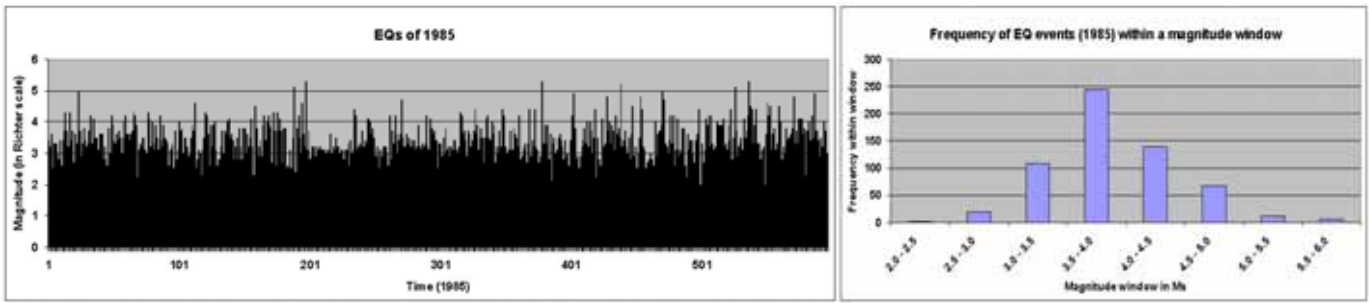

**Fig. 12. Magnitude as a function of occurrence time (left) and frequency distribution of magnitudes at 0.5R intervals (right) for the year 1985.**

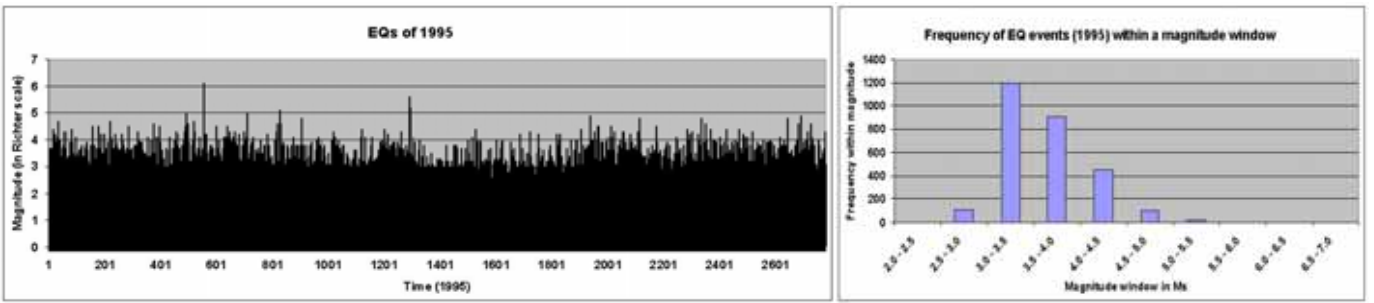

**Fig. 13. Magnitude as a function of occurrence time (left) and frequency distribution of magnitudes at 0.5R intervals (right) for the year 1995.**

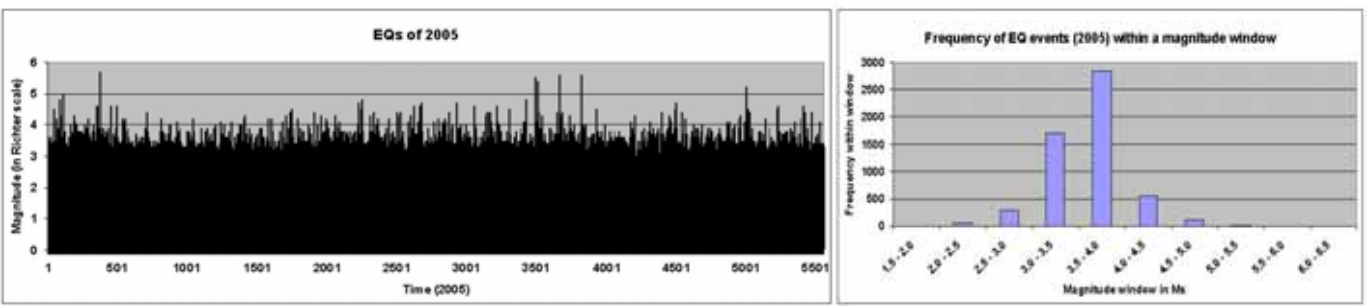

**Fig. 14. Magnitude as a function of occurrence time (left) and frequency distribution of magnitudes at 0.5R intervals (right) for the year 2005.**

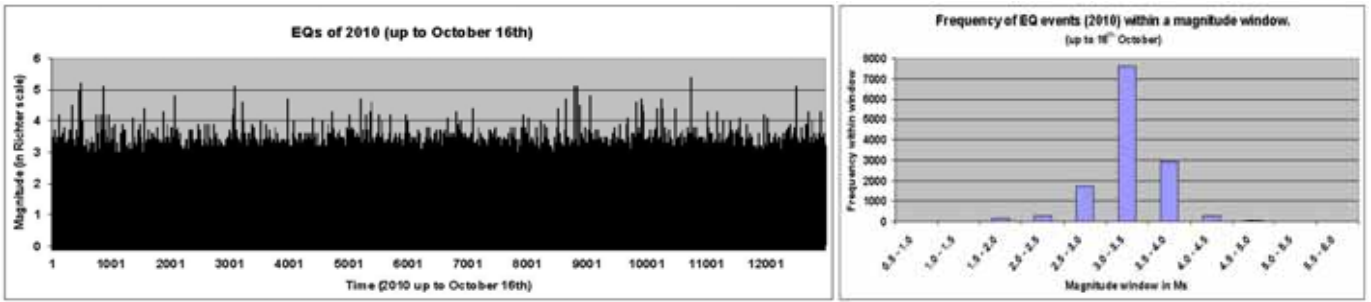

**Fig. 15. Magnitude as a function of occurrence time (left) and frequency distribution of magnitudes at 0.5R intervals (right) for the year 2010.**

A close inspection of figures (10) to (15) left, reveals that the most common seismic events are the ones with magnitude of the range of Ms = 3.5R. A better estimation is made when the distribution of the magnitude is drawn as an occurrence frequency plot (right). By taking into account only the magnitude window that corresponds to the largest occurrence frequency then the most frequent magnitude window is calculated as Ms = 3.41 – 3.92 R, by averaging the corresponding magnitude windows noted at the figures (10) to (15).

Moreover, left figures (10 – 15) show an exponential increase of the total seismic events per year that is shown in figure (16) right, while the same character is revealed by the most common in magnitude seismic events (fig. 16 left) of the magnitude window of Ms = 3.41 – 3.92 R.

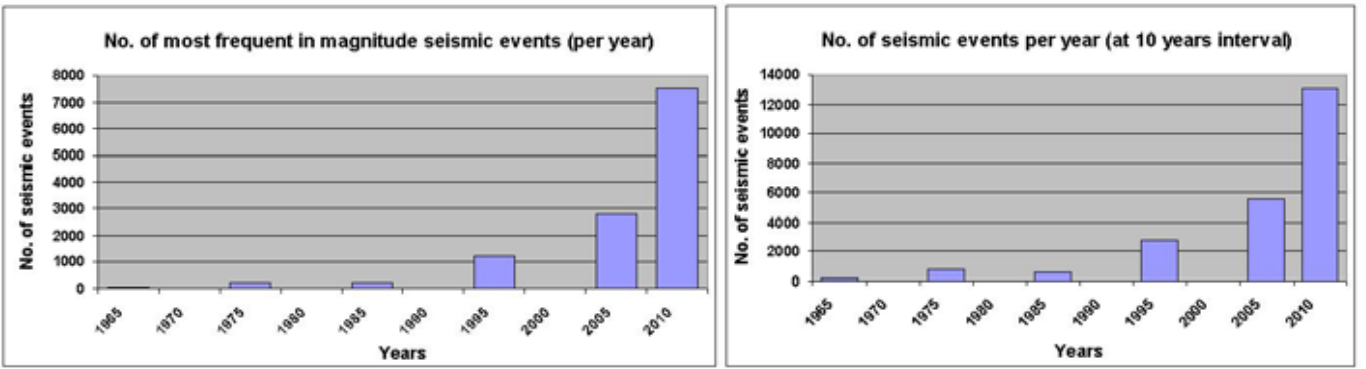

**Fig. 16. Most frequent magnitude seismic events (left) and total no. of seismic events per sample year (right).**

Therefore, a question is raised here on whether this observed increase is a real one caused by any tectonic cause or due to any change applied to the monitoring seismic network. A change in the seismic network recording sensitivity i.e. could easily produce similar results by increasing the no. of the recorded smaller magnitude seismic events. But this hypothesis should be rejected by the fact that "most common" seismic events (Ms = 3.41 – 3.92 R) were always detectable since 1964. Moreover, this exponential change has started almost 10 – 15 years ago while the monitoring seismic network is considered as being at rather stable performance conditions. Consequently, the observed seismicity increase can be attributed to the rather recently increasing large scale evolving tectonic events in the Greek territory.

In the following section the observed seismicity increase will be analyzed in terms of seismic energy release and discrete magnitude values of the occurred earthquakes for the period 1964 – 2010.

**3. Seismic data analysis.**

It has been shown that the cumulative seismic energy release of a seismogenic region can be represented by a linear function of time provided that the seismogenic area is under stable conditions, that is that the released seismic energy within a period of time equals with the one stored in it due to tectonic – stress processes (Thanassoulas 2007, 2008, 2008a, Thanassoulas et al. 2001). The stable conditions are expressed by the following equation (1) which holds for a certain time period while the cumulative seismic energy release is expressed by equation (2). Moreover, in the case when the parameter "C" of equation (2) is time depended then the well-known accelerating deformation results.

$$dE_{in} = dE_{out} \qquad (1)$$

$$E_{cum}(t) = C \cdot t + b \qquad (2)$$

$$E_{cum}(t) = k_{n+1} t^{n+1} + k_n t^n + \ldots\ldots k_0 \qquad (3)$$

An example of a cumulative seismic energy release graph is presented in the following figure (17). Periods of stable conditions (linear parts) as well as periods of accelerating deformation (non linear) can be distinguished very clearly in this graph. The magnitudes which were taken into account range from Ms = 3.0R to Ms = 7.5R. The considered seismogenic area is the entire Greek territory and the period of time spans from 1964 to 2010. The sampling interval (required by equation - 1) equals to one (1) month.

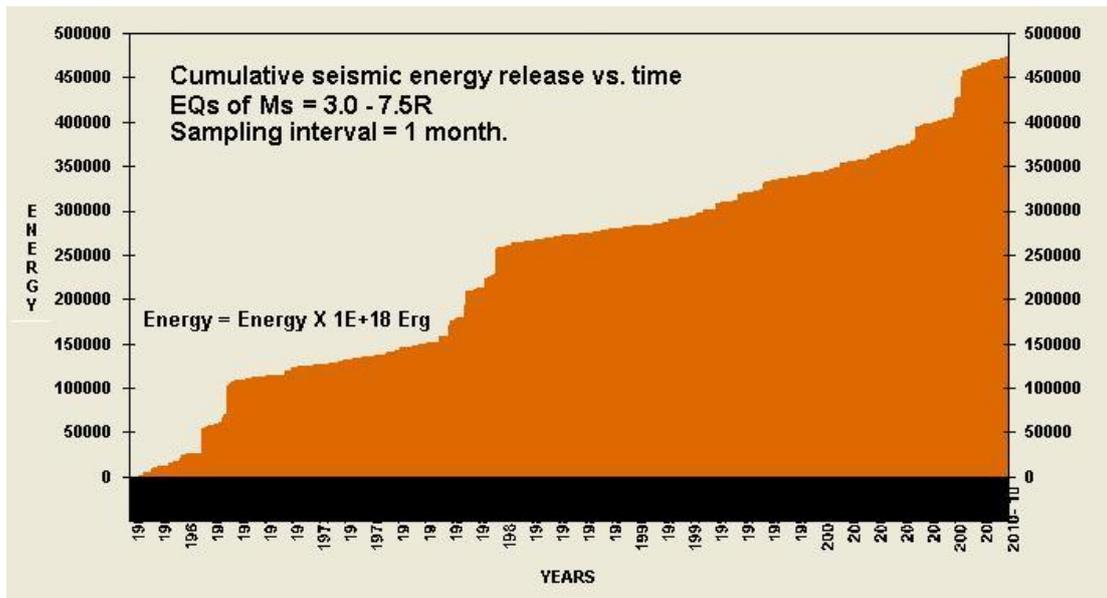

**Fig. 17. Cumulative seismic energy release graph of the Greek territory determined for the time period from 1964 to 2010.**

It is evident that the cumulative seismic energy release graph presented in figure (17) is composed by a general linear trend which is superimposed by some short wave length energy release anomalies. Typically, that seismic energy release time function could be decomposed into discrete frequency components (frequency spectrum) by any suitable FFT procedure. Even in this case, the discrete harmonic components will indicate only the oscillating character of the seismic energy release at each frequency but not any information about the magnitudes that participate at each frequency component.

A quite different approach for analyzing the seismic energy release graph is as follows. Since the graph of figure (17) is constructed by taking into account all the earthquakes, it is interesting to see how the seismic energy release behaves when only a unique magnitude of the occurred earthquakes is taken into account. This operation resembles the one of a sharp band-pass filter applied on the magnitudes which will be used for the construction of the seismic energy release graph. Therefore, by using different magnitude values, different cumulative seismic energy release graphs will be determined which will be the different components of the original total seismic energy release graph shown in figure (17).

Furthermore, by comparing the discrete seismic energy release graphs forms to equations (2, 3) it can be concluded at what earthquake magnitude there are stable tectonic – stress conditions, accelerating deformation or irregular occurrence of earthquakes.

At this point it is worth to re-postulate the "back-ground seismicity rate" as the one which exhibits, for a specific earthquake magnitude, linear cumulative seismic energy release. In this way it is possible to know, for a certain seismogenic area and for a specific sample time interval, what are the most common magnitudes of the earthquakes to occur.

The proposed analysis will be applied on the entire Greek territory, which will be considered as a single seismogenic area, and for magnitudes ranging from Ms = 3.0R - 7.5R at 0.5R intervals. The sampling time interval was chosen as 6 months in order to smooth the generated graphs from seismic noise produced by random seismic events. The latter is demonstrated by the following figures (18 – 22). These graphs have been generated for sampling time intervals starting from 10 days and progressively increasing to 15 days, 1 month, 3 months and 6 months. At each figure are presented the no. of the corresponding earthquakes for each time sample (left) and the corresponding cumulative seismic energy release graph (right).

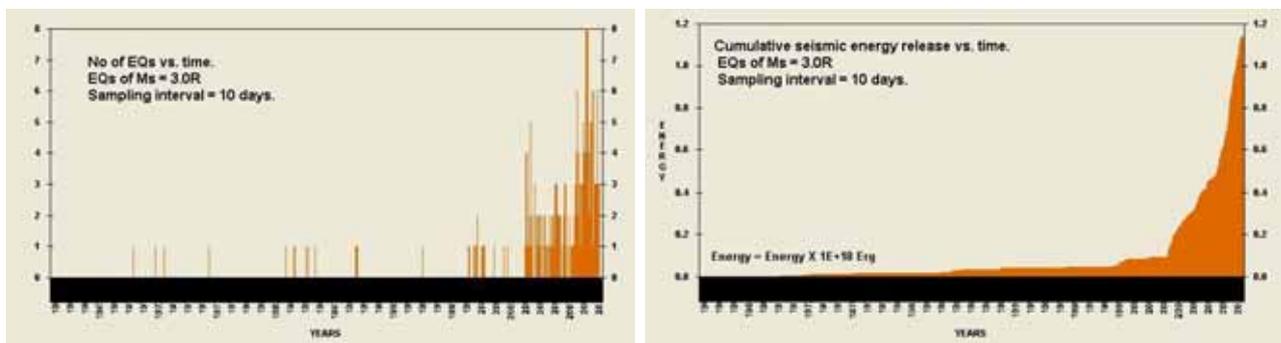

**Fig. 18. No. of earthquakes (left) and corresponding cumulative seismic energy release (right) as a function of time (1964 – 2010). Sampling interval = 10 days, Ms = 3.0R.**

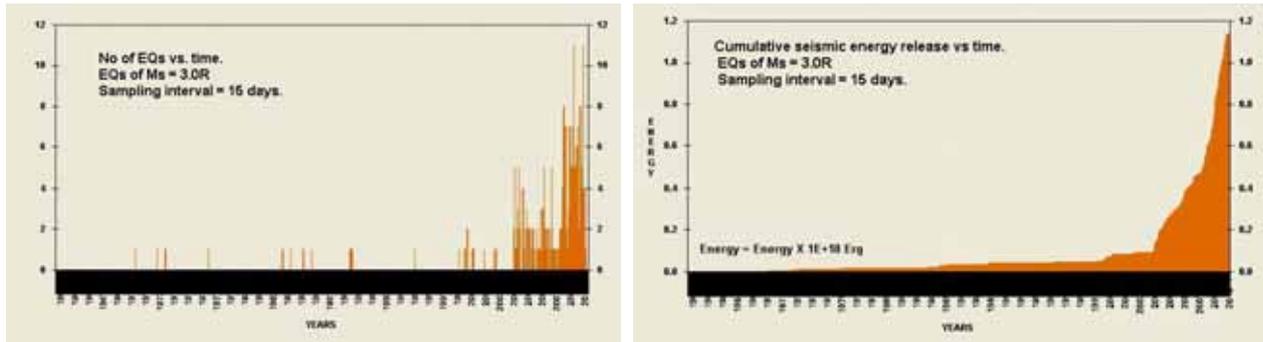

**Fig. 19. No. of earthquakes (left) and corresponding cumulative seismic energy release (right) as a function of time (1964 – 2010). Sampling interval = 15 days, Ms = 3.0R.**

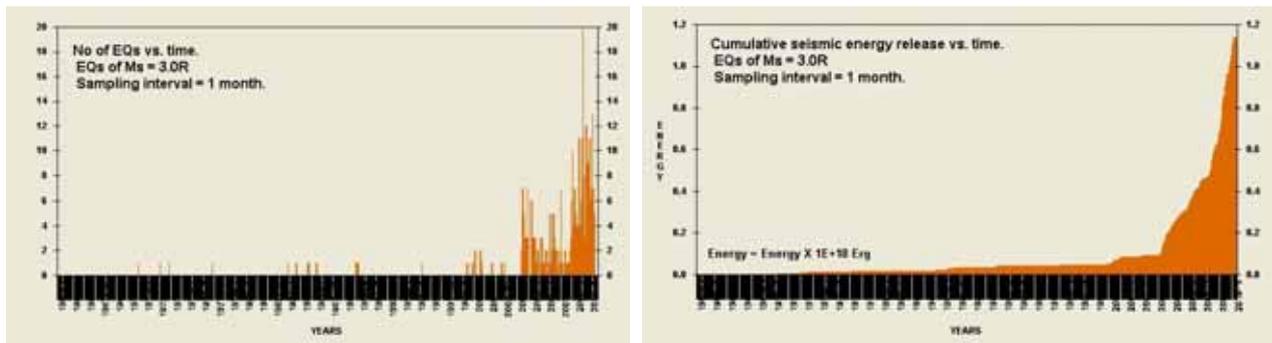

**Fig. 20. No. of earthquakes (left) and corresponding cumulative seismic energy release (right) as a function of time (1964 – 2010). Sampling interval = 1 month, Ms = 3.0R.**

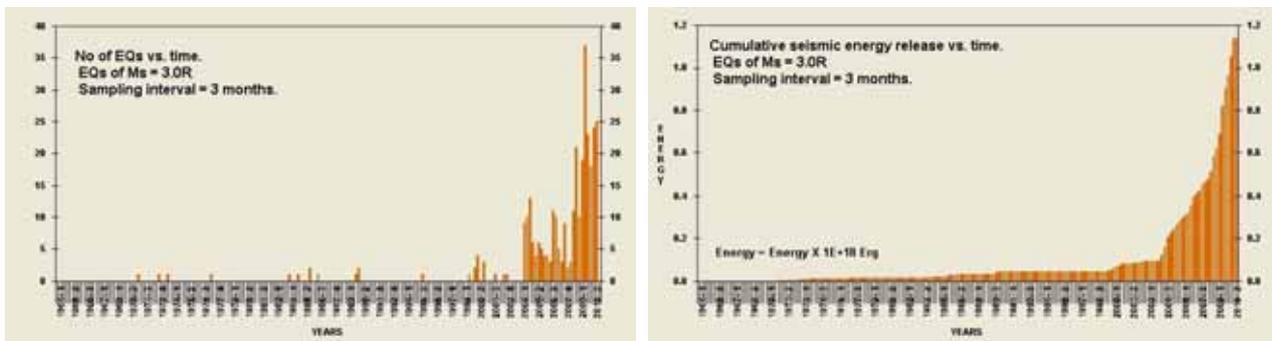

**Fig. 21. No. of earthquakes (left) and corresponding cumulative seismic energy release (right) as a function of time (1964 – 2010). Sampling interval = 3 months, Ms = 3.0R.**

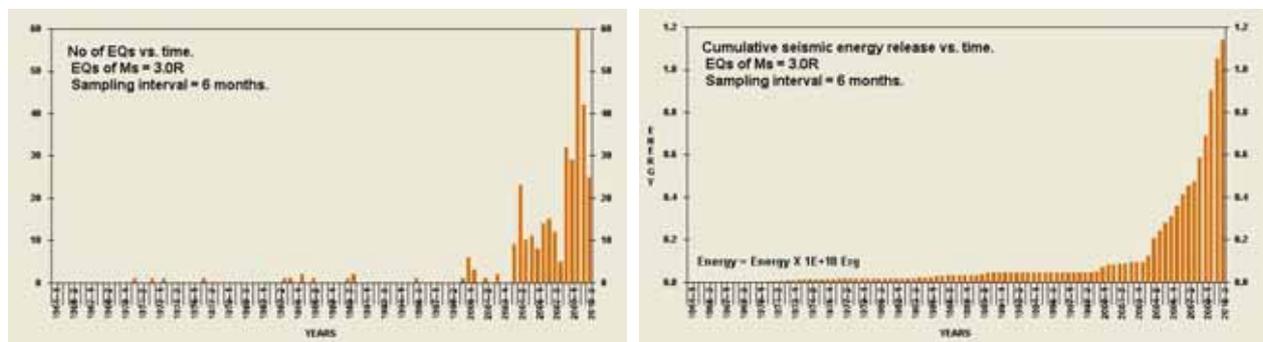

**Fig. 22. No. of earthquakes (left) and corresponding cumulative seismic energy release (right) as a function of time (1964 – 2010). Sampling interval = 6 months, Ms = 3.0R.**

It is evident from figures (18 – 22) that the overall graph character for the no. of earthquakes and cumulative seismic energy release does not change. Just the graphs become smoother as long as the sampling time interval increases. It is worth to notice the sharp increase of the last almost 10 years in both graphs (no. of earthquakes and cumulative seismic energy release).

In the following figures (23, 24, 25) which have been plotted for magnitudes Ms = 3.5R, 4.0R, 4.5R the increase of the no. of earthquakes and of the corresponding cumulative seismic energy release is still clearly present.

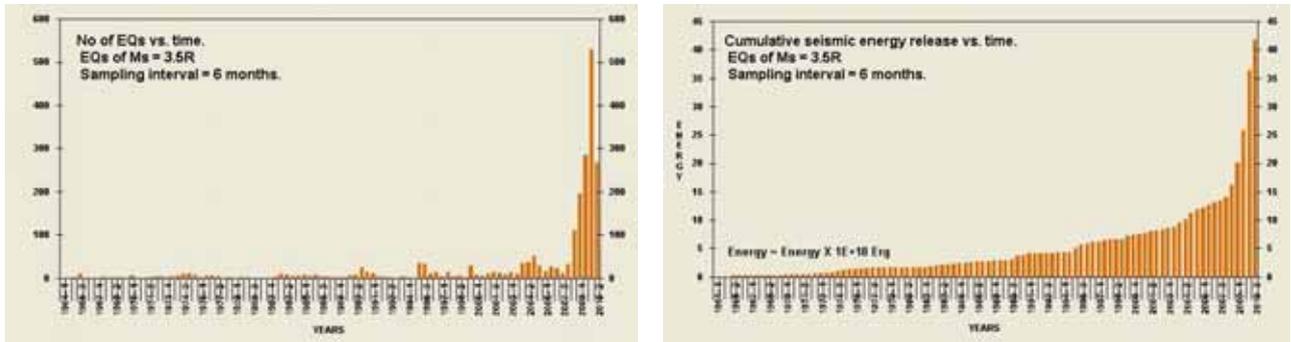

**Fig. 23. No. of earthquakes (left) and corresponding cumulative seismic energy release (right) as a function of time (1964 – 2010). Sampling interval = 6 months, Ms = 3.5R.**

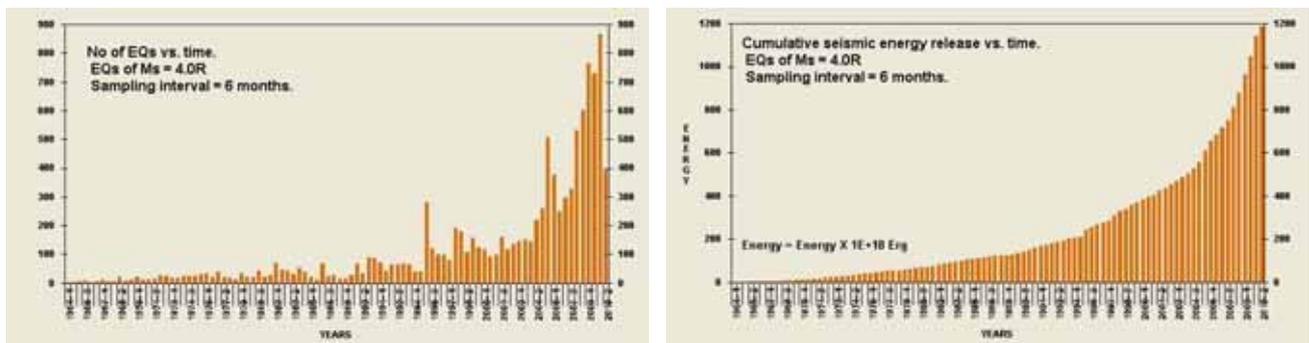

**Fig. 24. No. of earthquakes (left) and corresponding cumulative seismic energy release (right) as a function of time (1964 – 2010). Sampling interval = 6 months, Ms = 4.0R.**

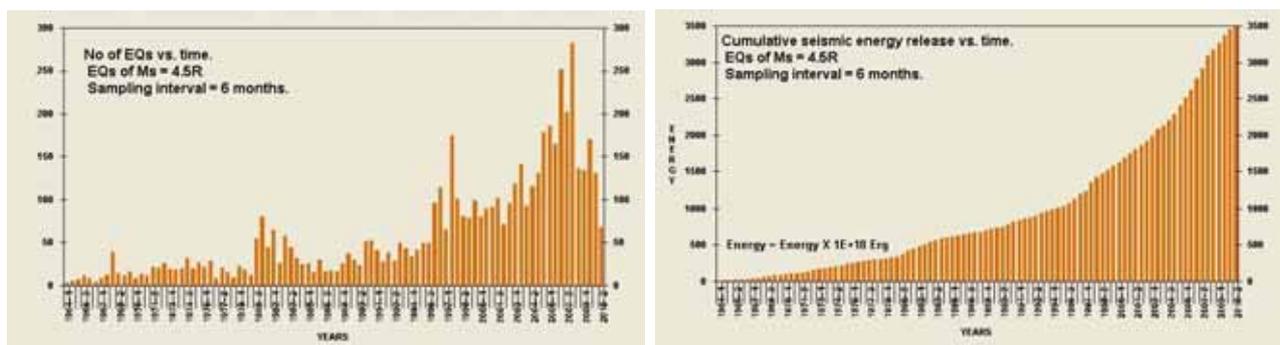

**Fig. 25. No. of earthquakes (left) and corresponding cumulative seismic energy release (right) as a function of time (1964 – 2010). Sampling interval = 6 months, Ms = 4.5R.**

What is suggested from figures 18 – 25 is that the entire Greek territory is at a state of accelerating deformation for magnitudes ranging from Ms = 3.0R to Ms = 4.5R.

The situation changes when larger magnitudes are considered. Following are shown the calculated graphs for magnitudes of Ms = 5.0R to Ms = 6.0R.

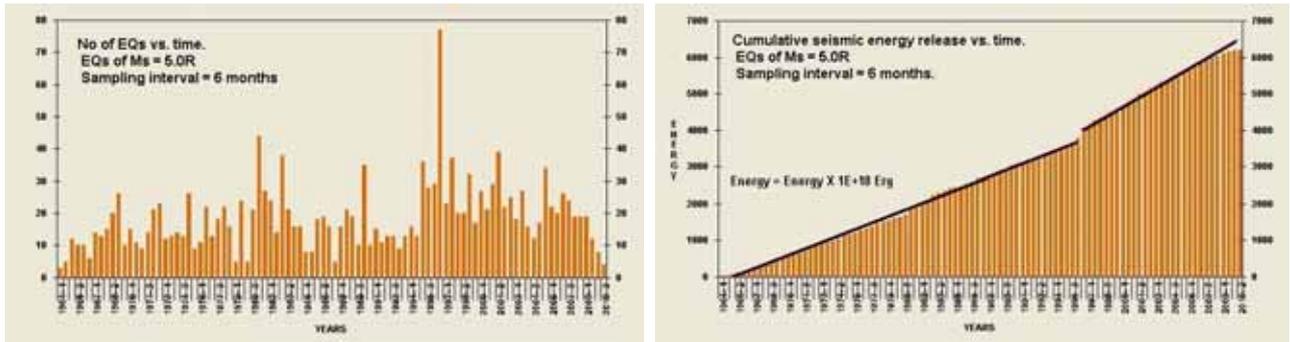

**Fig. 26. No. of earthquakes (left) and corresponding cumulative seismic energy release (right) as a function of time (1964 – 2010). Sampling interval = 6 months, Ms = 5.0R.**

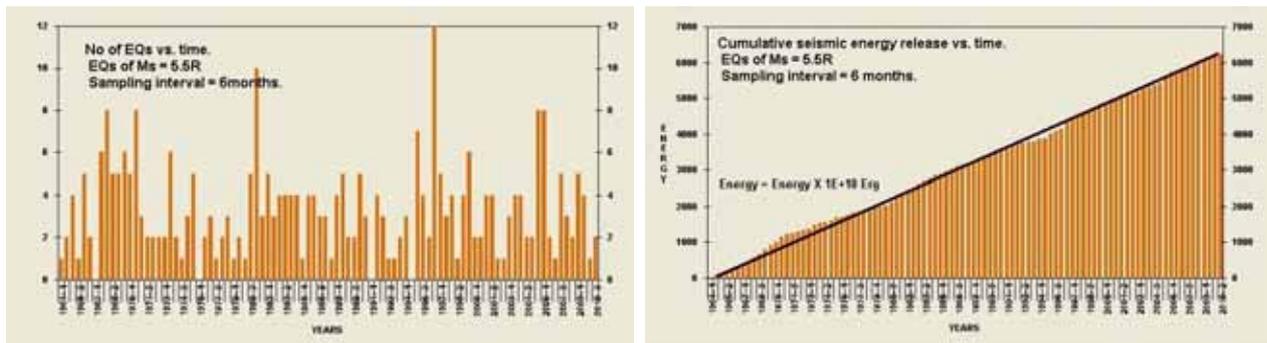

**Fig. 27. No. of earthquakes (left) and corresponding cumulative seismic energy release (right) as a function of time (1964 – 2010). Sampling interval = 6 months, Ms = 5.5R.**

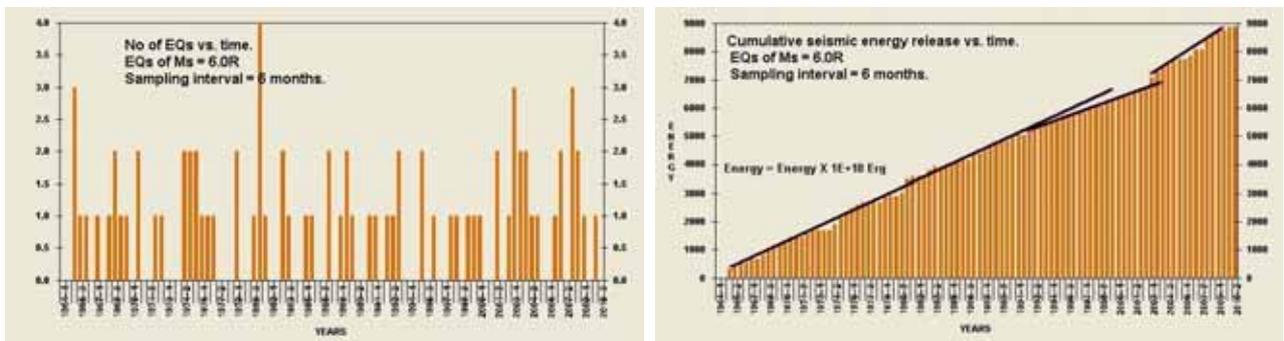

**Fig. 28. No. of earthquakes (left) and corresponding cumulative seismic energy release (right) as a function of time (1964 – 2010). Sampling interval = 6 months, Ms = 6.0R.**

**Figures (26, 27, 28) have changed in character compared with the ones (18 – 25) of smaller magnitudes. At first, there is no observable increase of the no. of earthquakes. The no. of earthquakes per time sample rather fluctuates in an irregular way around a mean value. The cumulative seismic energy release graph is characterized by large linear segments (black line segments) which suggest stable stress – tectonic conditions at these magnitude levels. From the latter graphs (26, 27, 28) only the figure (27) can be considered as representative for representing a stable linear condition which satisfies the equation (2). Therefore, the magnitude Ms = 5.5R can be considered as the normal "background seismicity rate" for the entire Greek territory for a sampling interval of six (6) months. In other words, for the study period, the no. of earthquakes per year with magnitude Ms = 5.5R must be constant. It is understood that small deviations from the calculated value of the "background seismic rate" will exist if a comparison is made between the actual no. of earthquakes of Ms = 5.5R of a specific year and the theoretical one which is the result determined through the LSQ method.**

**The last three figures (29, 30, 31) which have been constructed for the largest magnitudes (Ms = 6.5R – 7.5R) reveal the irregular character in time of the earthquake generation and consequently the corresponding irregular character of the cumulative seismic energy release.**

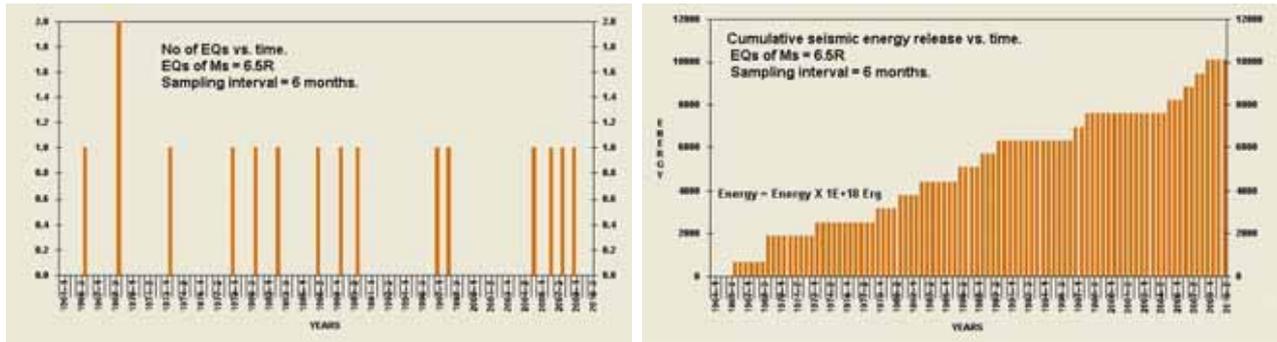

**Fig. 29.** No. of earthquakes (left) and corresponding cumulative seismic energy release (right) as a function of time (1964 – 2010). Sampling interval = 6 months, Ms = 6.5R.

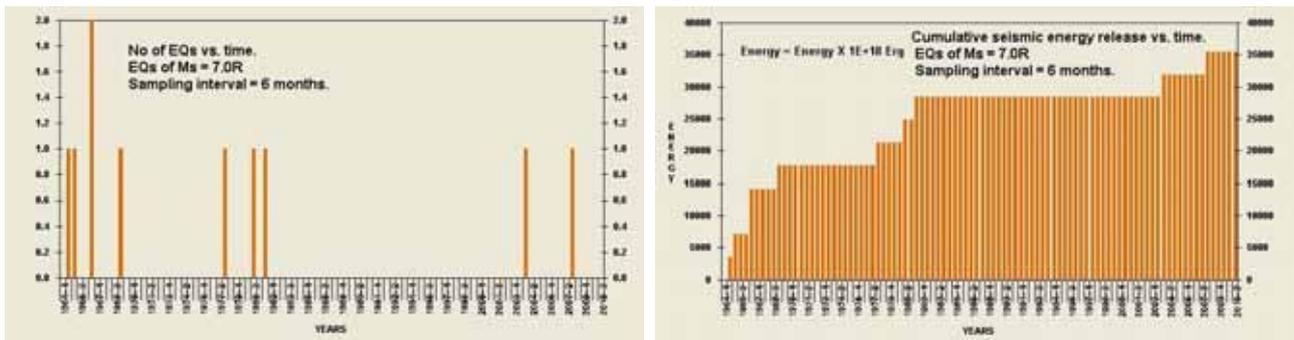

**Fig. 30.** No. of earthquakes (left) and corresponding cumulative seismic energy release (right) as a function of time (1964 – 2010). Sampling interval = 6 months, Ms = 7.0R.

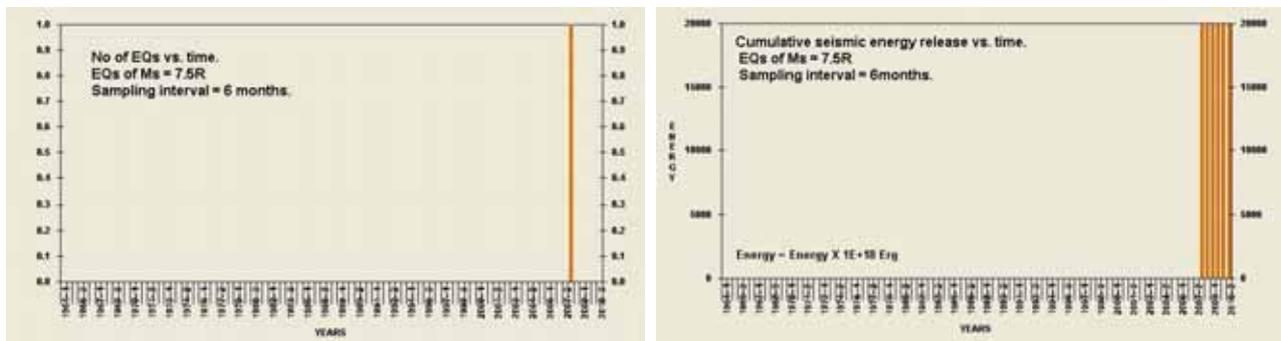

**Fig. 31.** No. of earthquakes (left) and corresponding cumulative seismic energy release (right) as a function of time (1964 – 2010). Sampling interval = 6 months, Ms = 7.5R.

## 4. Conclusions.

In this work it is introduced a new concept regarding the background seismicity rate. Instead of referring to the total seismicity, after having excluded the one due to clusters, as the background seismicity rate, it is used the total seismicity, clusters included, in terms of cumulative seismic energy release for a specific earthquake magnitude. It is understood that the seismogenic area and the time sampling period have been specified "a priori".

The analysis of the seismicity of the Greek territory, being considered as a unified seismogenic area, with the latter methodology, for the period of time from 1964 to 2010, revealed some interesting features.

- a. The Greek territory exhibits an accelerating no. of seismic events during almost the last 10 years. The magnitude Ms = 3.0R earthquakes spatial distribution indicates two regions of large seismic activity.
- b. Particularly, the observed increase is better expressed by earthquakes with magnitude of Ms = 3.0R to 4.5R. A better picture of this increase is given by the corresponding plot of the cumulative seismic energy release as a function of time. The observed increase of seismic events and consequently the corresponding cumulative seismic energy release, indicate that the Greek territory has entered a very long period of accelerating deformation. During the last few years, especially, the deformation is almost exponential in form.
- c. Earthquakes of magnitude of Ms = 5.0R to 6.0R fluctuate in no. irregularly around a mean value while the corresponding cumulative seismic energy release graphs mainly exhibit large in time linear segments. Particularly, the earthquakes of magnitude Ms = 5.5R generate a cumulative seismic energy release plot which fits nicely to a single straight line segment, for the entire study period, in an LSQ sense. Therefore, the magnitude of Ms = 5.5R can be considered as the normal "background seismicity rate" of the entire Greek territory, in terms of seismic energy release. Obviously it is quite larger than the one (Ms = 3.0R – 4.0R) adopted by the seismological community on the basis of the "most often occurred" in magnitude seismic events.
  d. Earthquakes of magnitude of Ms = 6.5R to 7.5R occur rather irregularly in time as it is shown by the corresponding graphs of no. of earthquakes occurrence and the corresponding cumulative seismic energy release.
  e. A detailed inspection of figure (26) generated for the earthquakes of magnitude Ms = 5.0R indicates that the cumulative seismic energy release graph consists of two segments. The first spans from 1964 to 1996 and the second spans from 1997 to 2010. Clearly, the second segment suggests increased no. of earthquakes per time unit during the last 13 years. Therefore, it could be said that accelerating deformation just starts to evolve even for earthquakes of magnitude of Ms = 5.0R.
  f. Earthquakes with magnitude larger than Ms = 5.5R exhibit a rather random occurrence time.
  g. All the previous observations can be related to the different phases that a seismogenic area undergoes before the occurrence of a large seismic event. These phases are shown in the following figure (32) as presented by Mjachkin et al (1975).

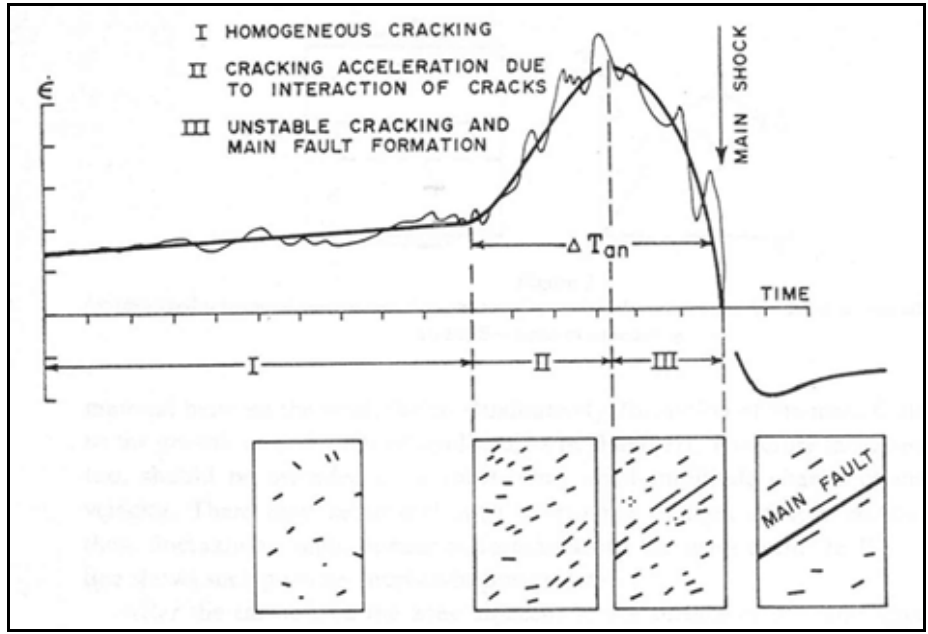

**Fig. 32. Different phases (after Mjachkin et al. 1975) that a seismogenic area undergoes before the occurrence of a large seismic event.**

In phase (I) homogeneous cracking is present with no accelerating mode observed. This is probably the state when only seismic noise (small magnitude seismic events) is registered. For the case of the Greek territory it is related to the earthquakes with magnitude of Ms = 3.0R which occurred during the period from 1964 to almost 2000 (fig. 22).

In phase (II) small cracks interact and acceleration is observed. Although only small seismic events take place acceleration is observed. This is demonstrated, for the case of the Greek territory, in figures (22 – 25) where acceleration is observed up to a magnitude of Ms = 4.5R.

In phase (III) unstable cracking and main fault formation is observed. Consequently, larger in magnitude earthquakes are observed. Figure (26) is a good example which indicates an increase of the generation of earthquakes with magnitude of Ms = 5.0R while no acceleration is observed for the magnitude of Ms = 5.5R indicated by the presence of a single linear segment on the cumulative seismic energy release graph.

In conclusion it can be said that the Greek territory, being considered as a unified seismogenic area, is at a state of accelerating deformation for the almost last 10 years for the earthquakes of magnitude up to Ms = 5.0R. The latter corresponds to phase (III) of the diagram of figure (32). Consequently, it is concluded that in the Greek territory a major seismic event is under preparation which will occur in the next years to come and most probably will be located at one of the two highly seismically active regions presented in figure (7, left).

## 5. References.

De Bremaecker, G., Huchon, P., LePichon, X., 1982. The deformation of Aegea: A finite element study., Tectonophysics, 86, pp. 197-211.

Gardner, J. K. and L. Knopoff (1974). Is the sequence of earthquakes in Southern California with aftershocks removed, Poissonian?, *Bull. Seism. Soc. Am.* 64, 1363-1367.

Habermann, R. E. (1983). Teleseismic detection in the Aleutian island arc, *J. Geophys. Res.* 88, 5056-5054.

HABERMANN, R. E., WYSS, M., 1984. BACKGROUND SEISMICITY RATES AND PRECURSORY SEISMIC QUIESCENCE: IMPERIAL VALLEY, CALIFORNIA. Bulletin of the Seismological Society of America, Vol. 74, No. 5, pp. 1743-1755.

Hadley, D. M. and D. S. Cavit (1982). A new methodology for identifying foreshocks, *EOS, Trans. Am. Geophys. Union* 63, 1042.

Hanus, V., and Vanek, J., 1993. Seismically active fracture zones related to the eastern segment of the Hellenic subduction, J. Geodynamics, Vol. 17, No. 1, pp. 39-56.

McNally, K. (1976). Spatial, temporal and mechanistic character in earthquake occurrence, *Ph.D. Thesis,* University of California, Berkeley, California.

Mjachkin, V.I., Brace, W., Sobolev, G.A., and Dieterich, J., 1975. Two models for earthquake forerunners, Pure Appl. Geophys. 113, pp. 169 -180.

NOA, Geodynamic Institute of Greece, National Observatory of Athens (NOA) seismic catalog: www.gein.noa.gr/services/

Papazachos, B.C., 1988. The seismic zones in the Aegean and surrounding area., "Europ. Seism. Com., XXI Gen. Assem., Sofia, Bulgaria, August 23-27, pp. 1-6.

Papazachos, B.C., and Kiratzi, A., 1996. A detailed study of the active crustal deformation in the Aegean and surrounding area., Tectonophysics, 253, pp. 129 – 153.

Savage, W. U. (1972). Microearthquake clustering near Fairview Peak, Nevada, and in the Nevada seismic zone, *J. Geophys. Res.* 77, 7049-7056.

Thanassoulas, C., 1998. Location of the seismically active zones of the Greek area by the study of the corresponding gravity field., Open File Report: A.3810, Inst. Geol. Min. Exploration (IGME), Athens, Greece, pp. 1-19.

Thanassoulas, C., 2007. Short-term Earthquake Prediction, H. Dounias & Co, Athens, Greece. ISBN No: 978-960-930268-5.

Thanassoulas, C. 2008. The seismogenic area in the lithosphere considered as an "Open Physical System". Its implications on some seismological aspects. Part – I. Accelerated deformation. arxiv.org/0806.4772 [physics.geo-ph]

Thanassoulas, C. 2008a. The seismogenic area in the lithosphere considered as an "Open Physical System". Its implications on some seismological aspects. Part – II. Maximum expected magnitude. arxiv.org/0807.0897 v1 [physics.geo-ph]

Thanassoulas, C., Klentos, V., 2001. The "energy-flow model" of the earth's lithosphere. Its application on the prediction of the "magnitude" of an imminent large earthquake. The "third paper". IGME, Open file report: A.4384, Greece.

Thanassoulas, C., Klentos, V. 2003. Seismic potential map of Greece, calculated by the application of the "Lithospheric energy flow model"., IGME, Open File Report A. 4402, Athens, Greece, pp. 1-25.

Thanassoulas, C., Klentos, V., 2008. The seismogenic area in the lithosphere considered as an "Open Physical System". Its implications on some seismological aspects. Part - III. Seismic Potential. arXiv:0807.1428. v1 [physics.geo-ph].

Thanassoulas, C., Klentos, V. 2010. Seismic potential map of Greece calculated for the years 2005 and 2010. Its correlation to the large (Ms>=6.0R) seismic events of the 2000 - 2009 period. arXiv:1001.1491 v1 [physics.geo-ph].

Wyss, M., and Toya, Y. 2000. Is Background Seismicity Produced at a Stationary Poissonian Rate?, Bulletin of the Seismological Society of America, 90, 5, pp. 1174–1187.